\documentclass[aps,prl,twocolumn,superscriptaddress]{revtex4-1}

\usepackage{graphicx}
\usepackage{amsmath}
\usepackage[dvips]{epsfig}
\usepackage[sort&compress]{natbib}
\usepackage{physics}
\usepackage{color}
\usepackage{ulem}

\newcommand\mymathop[1]{\mathop{\operatorname{#1}}} 

\begin{document}

\title{The effect of electron-phonon interactions on the spectral properties of single defects in hexagonal boron nitride}

\author{Ozan Ar{\i}}\email{ozanari@iyte.edu.tr}
\affiliation{Department of Physics,
Izmir Institute of Technology, Urla, Turkey}
\author{Nahit Polat}
\affiliation{Department of Physics,
Izmir Institute of Technology, Urla, Turkey}
\author{Volkan F{\i}rat}
\affiliation{Department of Physics,
Izmir Institute of Technology, Urla, Turkey}
\author{\"Ozg\"ur \c{C}ak{\i}r}
\affiliation{Department of Physics,
Izmir Institute of Technology, Urla, Turkey}
\author{Serkan Ates}\email{serkanates@iyte.edu.tr}
\affiliation{Department of Physics,
Izmir Institute of Technology, Urla, Turkey}
\date{\today}

\date{\today}

\begin{abstract}

We investigate temperature-dependent spectral properties of a single defect in hexagonal boron nitride (hBN). We observe a sharp zero-phonon line (ZPL) emission accompanied by Stokes and anti-Stokes optical phonon sidebands assisted by the Raman active low-energy ($\approx~6.5$~meV) interlayer shear mode of hBN. Spectral lineshape around the ZPL is measured down to 78~K, at which the linewidth of the ZPL is measured as 172~$\mu$eV. By employing a quadratic electron-phonon interaction, the temperature-dependent broadening and the lineshift of the ZPL are found to follow $T+T^5$ and $T+T^3$ temperature dependence, respectively. Furthermore, the temperature-dependent lineshape around the ZPL is modeled with a linear electron-phonon coupling theory, which results in the Debye-Waller factor of the ZPL emission as 0.59.
\end{abstract}



\maketitle

Understanding the basic optical properties of isolated quantum light sources and controlling their emission properties allow them to be used effectively in many areas within quantum information technologies. In literature, semiconductor quantum dots, single molecules, and nitrogen vacancy centers in diamond are widely studied systems among a large number of known single photon sources \cite{Aharonovich2016}. In particular, isolated color centers in three-dimensional wide bandgap materials (i.e., diamond, silicon carbide, zinc oxide) are of great interest as single-photon sources. Recently, efficient single-photon generation from two-dimensional (2D) transition metal dichalcogenides (TMDCs) and defects in hexagonal boron nitride (hBN) are demonstrated\cite{He2015,Srivastava2015a,Chakraborty2015,Tran2015}. Each of these materials has their own advantages in terms of operating conditions such as working temperature or radiation energy, which can be important for different applications.

In particular, defects in hBN have attracted a great interest due to their unique properties. The quantum nature of the emission obtained from defects in this material is not limited to only cryogenic temperatures, which indeed can be preserved for up to 800~K \cite{Kianinia2017}. In addition, because of its large bandgap ($\approx 6$~eV), hBN can host several types of defects which emit over a large spectral range \citep{Jungwirth2016}. Finally, due to its two-dimensional nature, optically-active defects in hBN can be very close to the sample surface and these defects can interact very efficiently with other two-dimensional structures and photonic devices \cite{Kim2018b}. Despite the above-mentioned advantageous, zero-phonon line (ZPL) emission from a defect in hBN is strongly affected by the vibrational properties of the crystal \cite{Vuong2016,Jungwirth2016,Jungwirth2017,Exarhos2017}, which influences the performance of these sources greatly for the applications. 

In this letter, we present a quantitative study on spectral properties of an optically active single defect in hBN. A sharp ZPL and optical phonon sidebands due to the low-frequency interlayer shear mode of bulk hBN are observed for the first time in an emission spectrum. Temperature-dependent micro-photoluminescence ($\mu$-PL)  spectra are compared with a theoretical model which is based on electron-phonon interactions including both the linear and the quadratic displacement terms. By taking into account that acoustic phonons coupled to the electronic states of the defect through the deformation potential and the piezoelectric coupling, we show that the spectral properties of ZPL emission are strongly influenced by the vibrational properties of hBN.  An excellent agreement between the experimental results and theoretical calculations reveals that the relevant theoretical model can be used for accurate calculation of the Debye-Waller and the Huang-Rhys factors, both of which are commonly used to determine the potential of single quantum emitters. 

\begin{figure}[t]
\centerline{\includegraphics[width= 0.48\textwidth]{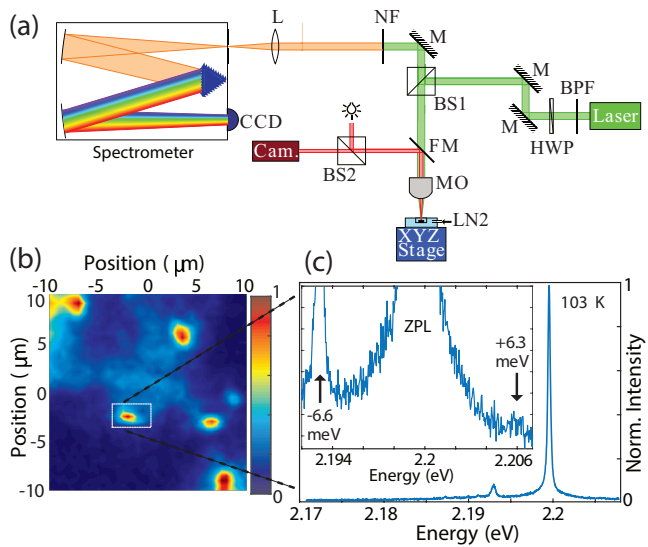}}
\caption{(a) Schematic of the confocal micro-PL setup, where BS1 and BS2 are beam splitters with 30:70 (R:T) and 10:90 (R:T) ratio, respectively.  (b) Normalized PL map of the studied bulk hBN structure, which shows bright and localized emission from several defects. (c) PL spectrum of the selected defect with a ZPL emission near 2.2~eV at 103~K. The inset shows zoomed-in plot around ZPL, where phonon-assisted Stokes and anti-Stokes transitions are observed at -6.6~meV and 6.3~meV, respectively.} 
\label{fig:Fig1}
\end{figure}

In this work, multilayer hBN flakes (obtained in a solution from Graphene Supermarket) drop-casted on silicon dioxide substrate are used as the material system. Optical properties of the sample are investigated using a confocal micro-PL system as shown in Fig. \ref{fig:Fig1} (a). The sample is placed in a liquid nitrogen cryostat and mounted on a high precision XYZ translation stage.  A microscope objective (NA~=~0.5) is used to focus the excitation source (Argon laser at 488~nm) on the sample and to collect the emission, which is later dispersed on a monochromator equipped with an 1800~grooves/mm grating and an EMCCD camera (spectral resolution $\approx 120~\mu$eV at the studied wavelength). Figure~\ref{fig:Fig1} (b) shows a spatially resolved normalized PL map of a multilayer hBN structure. Five isolated defects ($\approx1.2~\mu$m in size, limited by the spot size) with different ZPL energies are clearly visible. All results reported here are obtained from the selected defect that has an asymmetric ZPL emission around 2.2~eV (563~nm) at 103~K as shown in Fig.~\ref{fig:Fig1} (c). The linewidth of the peak is about  $200~\mu$eV, which is close to the resolution of the system. The spectrum also shows a small peak on the lower energy side of the ZPL with a detuning of $-6.6$~meV. 

To understand the origin of the observed spectral features, such as the position of the peaks with respect to each other and the asymmetric lineshape of the ZPL, the interaction between the excited electron of the defect and the vibrational properties of the host crystal needs to be considered. Phonon dispersion of hBN shows three acoustic phonons with energies up to $\approx 140$~meV and 9 zone-center optical phonons~\cite{Geick1966,Kuzuba1978,Nemanich1981,Cusco2016}. Four of these optical phonons are LO-TO degenerate Raman-active ($E_{2g}$) with energies reported as $\approx 6.5$~meV for interlayer shear mode and $\approx 169$~meV for in-plane mode in bulk structures at room temperature. The other optical phonons are either infrared-active or silent, which can not be observed in Raman spectra of hBN. A typical PL spectrum of an optically active defect in hBN, therefore, consists of a ZPL emission broadened by the interaction with acoustic phonons and phonon sidebands at certain energies assisted by the emission of optical phonons.  A phonon sideband due to the emission of high-energy in-plane $E_{2g}$ phonons is observed in the PL spectrum of a single defect in hBN \cite{Tran2015,Martinez2016}. Low-energy $E_{2g}$ optical phonon sideband coupled to an indirect excitonic emission of bulk hBN was observed \cite{Vuong2017a}, but has not been reported n a PL spectrum of a defect yet mostly because of its very weak interlayer interaction with electronic states of a defect that is well confined to a small area in a single layer. In addition, this mode has a much lower polarizability \cite{Cusco2016} with respect to the high-energy Raman-active phonon mode as confirmed by the large ratio of their intensities (a factor of $\approx 50$) observed in Raman signal of bulk hBN \cite{Kuzuba1978,Nemanich1981,Stenger2017}.  

The peak on the lower energy side of the ZPL with a detuning of $-6.6$~meV is considered to be the phonon sideband  emission from the defect (Stokes emission). It is assisted by the emission of low energy Raman active phonon mode of hBN crystal at the $\Gamma$ point. As the thermal energy $k_BT~=~8.8$~meV at 103~K is larger than the energy of the phonon mode, a weaker sideband on the higher energy side of the ZPL  is expected due to the absorption of this phonon (anti-Stokes emission). As shown in the inset of Fig.~\ref{fig:Fig1} (c), a small peak with a detuning of $\approx 6.3$~meV is observed on the higher energy side of the ZPL. The slight asymmetry between the energies of the Stokes and the anti-Stokes peaks might arise from different coupling strengths of ground and excited states of the defect to the phonon mode \cite{Norambuena2016}. To the best of our knowledge, this is the first observation of low energy Raman-active phonon mode in a PL spectrum of a defect in hBN.  

\begin{figure*}[t]
\centerline{\includegraphics[width=0.9\textwidth]{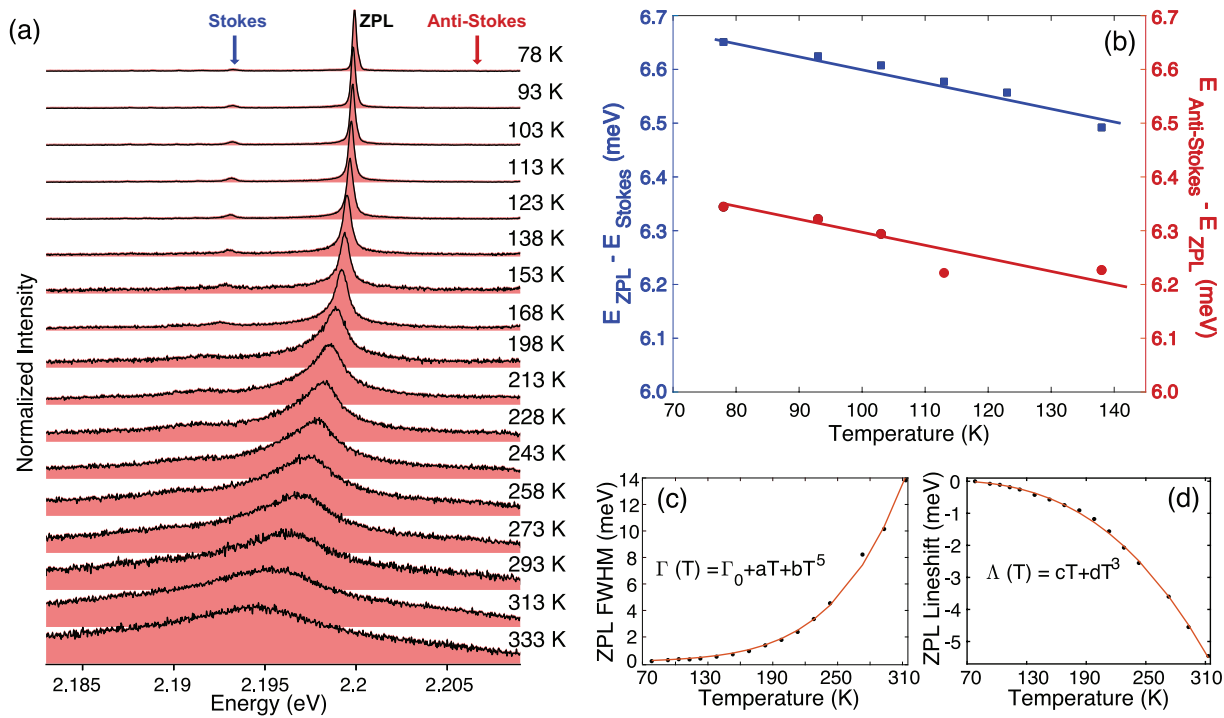}}
\caption{(a) Temperature-dependent $\mu$-PL spectra of the defect taken under same excitation power and polarization conditions. Each spectrum is normalized and shifted vertically for clarity. Evolution of the phonon assisted Stokes emission due to low energy Raman active phonon mode is observed as the temperature is reduced. The weak anti-Stokes emission due to the absorption of the same phonon mode is only visible in zoomed-in plot shown in Fig.~\ref{fig:Fig1} (c). (b) Temperature-dependent energy difference of ZPL-Stokes (left) and anti-Stokes-ZPL emission. (c) Temperature-dependent linewidth (c) and lineshift (d) of the ZPL. The solid lines are fit to the experimental data obtained from the model.}
\label{fig:Fig2}
\end{figure*}

To quantify the influence of electron-phonon interactions on the spectral properties of the observed emission, we performed systematic temperature-dependent micro-PL measurements on the selected defect under same excitation power and polarization conditions. Figure~\ref{fig:Fig2} (a) shows the normalized PL spectra of the defect taken at different  temperatures from room temperature to 78~K.  While the room temperature spectrum is mainly dominated by a broad asymmetrical peak at 2.195 eV, two distinct emission lines appear (Stokes and anti-Stokes phonon sidebands as discussed before) at the lower temperatures with a strongly narrowed and energetically shifted ZPL emission. As seen in  Fig.~\ref{fig:Fig2}(b), the detuning between the sidebands and the ZPL emission is reduced almost linearly as the temperature is increased (See supplementary information). This can be explained by the positive thermal expansion coefficient of hBN crystal along c-axis \cite{Paszkowicz2002}, which increases the interlayer spacing and, therefore, reduces the energy of the low-energy $E_{2g}$ phonon mode. A very similar temperature-dependent behaviour of this mode is reported in Raman scattering experiments \cite{Cusco2016,Vuong2017a,Stenger2017} but has not been observed in an emission spectrum of a single defect in hBN before. 

In addition to the properties of the optical phonon sidebands, the effect of temperature on the spectral properties of the ZPL emission is also investigated. Figure~\ref{fig:Fig2} (c) shows the temperature-dependent linewidth with a strong narrowing at the lower temperatures. Due to the spectral resolution of the micro-PL setup, the FWHM values below 100~K are deconvoluted with the instrument response function. The linewidth at 78~K is extracted as 172~$\mu$eV, which is even narrower than the reported values at 5~K for defects in hBN~\cite{Jungwirth2016} and defects in diamond (i.e., silicon or chromium vacancies) \cite{Muller2012,Neu2013}. However, it is still broader than the temperature-independent natural linewidth of a typical ZPL emission from a defect in hBN reported as $<$0.35~$\mu$eV~\cite{Jungwirth2016,Dietrich2018}.

At finite temperatures, the spectral diffusion and interaction with acoustic phonons broaden the linewidth of the ZPL emission from quantum emitters in solids \cite{Neu2013,Muller2012,Wolters2013,Fu2009}. Temperature-dependent linewidth of the ZPL emission from semiconductor quantum dots \cite{Favero2003,Besombes2001} and color centers in diamond \cite{Hizhnyakov2002,Neu2013} were studied using theoretical models relying on interaction with acoustic phonons. Recently, the linewidth of ZPL from defects in hBN were investigated by an effective single frequency phonon-mediated mechanism \cite{Jungwirth2016} or by employing excitonic states for emission near band gap of hBN \cite{Vuong2016}.  Here, the temperature-dependent behavior of ZPL linewidth is studied with a phonon-assisted broadening mechanism \textit{with continuum of phonon modes} interacting with the deep and isolated electronic states of a single defect. The broadening of the ZPL is considered through the phonon processes involving linear and quadratic electron-phonon coupling terms which is valid for systems exhibiting weak interaction with acoustic phonons \cite{Davies1974}. Considering the phonon density of states in 2D ($D(\omega) \propto \omega$) and the quadratic electron-phonon coupling, acoustic phonon modes introduce a broadening to the ZPL emission via the  piezoelectric coupling ($V(\omega) \propto 1/\sqrt{\hbar \omega }$) proportional to $T$ and the deformation potential ($V(\omega) \propto \omega/\sqrt{\omega }$) with $T^5$ (See ref. \cite{Stonehamm2001} and supplementary information for details). As the emission spectra ere taken well below the Debye temperature \cite{Tohei2006}, a constant contribution from the the linear electron-phonon coupling to broadening \cite{Stonehamm2001} is also expected. Therefore, the temperature-dependent linewidth of ZPL emission from a defect in hBN is obtained as;
\begin{equation}
\Gamma(T)=\Gamma_0+aT+bT^5 \label{eq:gamma}
\end{equation}
where the first term has the contribution from both the natural linewidth of the ZPL and the linear electron-phonon coupling while the second and the third terms represent the contributions from the quadratic electron phonon interactions via the piezoelectric coupling and the deformation potential, respectively. Figure~\ref{fig:Fig2} (c) shows the experimental linewidth extracted from each spectrum and the fit obtained from Eqn.~\ref{eq:gamma}, which results in $\Gamma_0$ of the emission as 0.83 $\mu$eV, similar to the reported values for the natural linewidth of defects in hBN~\cite{Jungwirth2016,Dietrich2018}. 

The energy of the ZPL has also a strong temperature dependence as observed on the emission spectra. Similar behavior is observed for different quantum emitters in solids, such as semiconductor quantum dots \cite{Favero2003,Besombes2001}  and defects in diamond \cite{Hizhnyakov2002,Muller2012,Neu2013}. Recently, a red shift of ZPL emission from  defects in hBN were reported, the reason for which were considered to be the change in bulk lattice constant \cite{Jungwirth2016} or fluctuating fields due to nearby defects \cite{Sontheimer2017}. In order to understand the red shift observed in Fig.~\ref{fig:Fig2}(a), we consider the coupling to the LA phonons through the deformation potential and the piezoelectric coupling with the density of states in 2D as before. The lowest non-vanishing contribution from the electron-phonon interaction to the lineshift is the quadratic coupling term \cite{Stonehamm2001}, which is calculated as proportional to $T$ and $T^3$ for the piezoelectric coupling and the deformation potential, respectively. The solid line in Fig.~\ref{fig:Fig2} (d) represents the fit to the experimental data with the  function given as $\Lambda(T) =cT+d T^3$. As seen, the phonon-mediated mechanism describes the lineshift behavior  of the emission from a single defect in hBN quite well (see supplementary information for the details).

Finally, the lineshape of the ZPL emission at a finite temperature is modeled with the linear electron-phonon coupling theory, which is shown to be valid as long as the electron-phonon interaction is weak \cite{Keil1965,Davies1974} at low temperature. Phonon assisted modeling of the lineshape was widely used to study the emission spectrum from quantum dots \cite{Krummheuer2002,Favero2003,Besombes2001}, defect centers in diamond \cite{Davies1974,Alkauskas2014} and more recently point defects in hBN \cite{Exarhos2017}. Here, the lineshape is calculated up to two-phonon proces (O($S^2$)) with ZPL, one-phonon line (OPL), and two-phonon line (TPL). The lineshape function is given as an overlap of the initial and final vibrational states \cite{Stonehamm2001}:
\begin{equation}
G(\omega)=\mymathop{Av}_{n} \sum_{n'} \prod_\alpha \left| \sigma(n_\alpha, n'_\alpha)\right|^2 \delta(\omega-\omega_{ij,nn'})
\label{eq:Lineshape_1}
\end{equation} 
where $n^{'}_\alpha$ is the occupation number of the phonon mode $\alpha$. The frequency of the emitted photon is given by 
$\hbar\omega_{ij,nn'}=E_{in}-E_{jn'}$ which is the energy difference between the excited electronic state $i$ in phonon state $\{n_\alpha\}$  and the ground electronic state $j$ in phonon 
state $\{n_\alpha'\}$. In linear coupling model this energy difference is equal to $\hbar\omega_{ij,nn'}=\hbar\omega_{ZPL}+(n-n')\hbar \omega$, 
where $\hbar\omega_{ZPL}$ is 
zero phonon line energy. Here also a thermal averaging is performed over the initial phonon states. 
The lattice oscillator overlaps for different modes can be expressed in terms of Huang-Rhys factors $S_\alpha$  as \cite{Keil1965}:
\begin{equation}
\left| \sigma(n_\alpha, m_\alpha)\right|^2 = S^{n_\alpha-m_\alpha} e^{-S_\alpha} \frac{m_\alpha !}{n_\alpha !} \left\lbrace L^{n_\alpha-m_\alpha}_{m_\alpha} 
(S_\alpha)\right\rbrace^2 .
\end{equation}
where $S_\alpha=(V(\omega_\alpha)/\hbar\omega_{\alpha})^2$.

Finally, the lineshape function defined in Eqn.(\ref{eq:Lineshape_1}) is obtained as:
\begin{equation}
\begin{split}
G(\omega) =\mymathop{Av}_{n_\alpha}  \sum_{m_\alpha} \prod_\alpha S_\alpha^{n_\alpha-m_\alpha} \: e^{-S_\alpha} \frac{m_\alpha !}{n_\alpha !} \\
\times \left\lbrace L^{n_\alpha-m_\alpha}_{m_\alpha} (S_\alpha)\right\rbrace^2 \delta(\omega-\omega_{ij,n_\alpha m_\alpha}).
\end{split}
\end{equation}
From the above expression, the ZPL is calculated to the second order in Huang-Rhys factor as follows:
\begin{equation}
\begin{split}
& G_0 (\omega) \simeq  \left[1-  \sum_\alpha  (2 \expval{n_\alpha}+1) S_\alpha + \sum_\alpha \expval{n_\alpha}^2 S_\alpha^2 \right. \\ &+ \left. \sum_{\alpha,\beta}  \expval{n_\alpha} S_\alpha  \expval{n_\beta} S_\beta + \sum_{\alpha,\beta} S_\alpha  (2 \expval{n_\beta} S_\beta) \right] \\ &\times \delta(\omega-\omega_{ij}) 
\end{split}
\end{equation}
where $S(\omega)$ is the energy dependent Huang-Rhys factor and $n(\omega)$ is the Bose-Einstein distribution function. Similarly, emission part of the OPL is given as;
\begin{equation}
\begin{split}
& G_1 (\omega)  =\mymathop{Av}_{n_\alpha, n_\beta} \left\lbrace (1-\sum_\alpha S_\alpha) \sum_\alpha (\expval{n_\alpha}+1)  S_\alpha \right. \\ &+ \left. \sum_{\alpha,\beta} (\expval{n_\alpha}+1) S_\alpha (-2 \expval{n_\beta}) S_\beta  - \sum_\alpha (\expval{n_\alpha}+1) \expval{n_\alpha}  S_\alpha^2 \right\rbrace \\ &\times \delta(\omega-\omega_{ij}-\omega_\alpha)  .
\end{split}
\end{equation}

\begin{figure}[t]
\centerline{\includegraphics[width=8 cm]{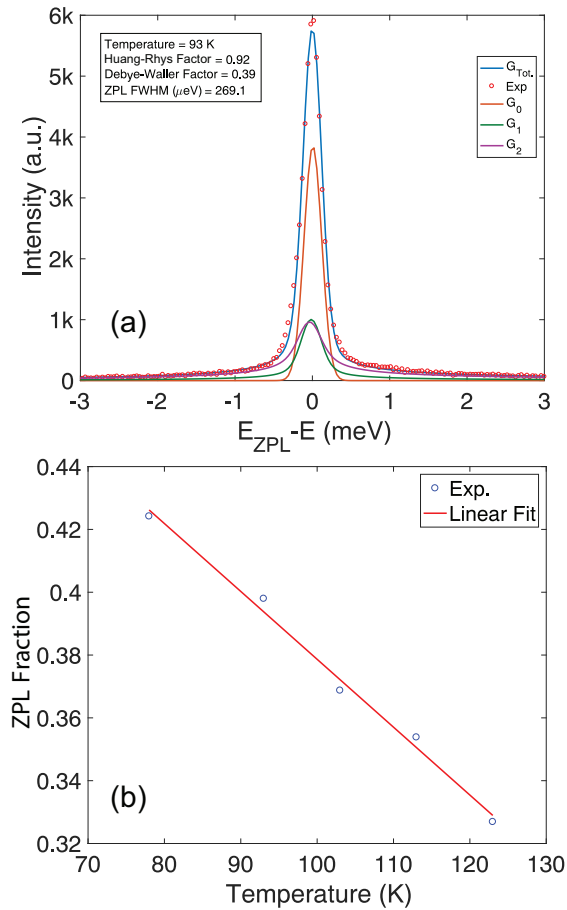}}
\caption{(a) ZPL emission of the defect at 93~K (symbols). The blue solid line is the lineshape calculated from the linear electron coupling theory with the contributions from ZPL (orange line), OPL (green line), and TPL (red line). (b) Temperature-dependence of the ZPL fraction estimated from the lineshape analysis by the linear electron-phonon coupling theory (symbols). The solid line is the linear fit to ZPL fractions. Huang-Rhys and Debye-Waller factors are calculated from the fit as 0.52 and 0.59 at 0~K, respectively.} \label{fig:Fig3}
\end{figure}

The absorption part of the OPL is calculated by replacing $n(\omega)$ terms with $(n(\omega)-1)$ in $G_1$ function (see Figure 1 in supplementary information for details). Finally, the TPL function ($G_2$) is obtained from the convolution of the OPL with only O($S$) terms. 

To produce a lineshape that fits to the measured spectrum, the discrete phonon peaks are replaced with the convolution of the ZPL ($G_0$) lineshape. Additionally, the lineshape is calculated up to a maximum phonon energy of $k_BT$, where $k_B$ is the Boltzman constant and $T$ is the temperature at which the data is taken. The Huang-Rhys terms for the phonons with the frequency of $\omega_\alpha$ is given as $S_\alpha\propto 1/\omega_\alpha$ for  the deformation potential while it is $S_\alpha\propto 1/\omega_\alpha^3$ for the piezoelectric coupling. The effect of the piezoelectric coupling is observed to be weak compared to the deformation potential and therefore it is ignored. Figure~\ref{fig:Fig3} (a) shows the experimental spectrum (symbols) obtained at 93~K and the results of the fit with the theoretical lineshape including up to two phonon processes with O($S^2$) and  the phonon sidebands from LA phonons. As seen from the figure, the dominant contribution to the lineshape is from the ZPL while the OPL and the TPL contribute mainly to the tails of the emission spectrum. The fraction of the ZPL (known as the Debye-Waller factor) within the total emission lineshape is calculated approximately by using $G_0 / (G_0+G_1 + G_2)$, where the terms in the denominator have contributions from the phonon-sidebands due to LA phonons. We note that, at low temperatures ($<$124~K) the occupation of the optical phonons \cite{Cusco2016} is negligible and the weak contribution from the sidebands are ignored for the calculation of the ZPL fraction. Therefore, a ZPL fraction of 0.39 is calculated at 93~K. To determine the temperature dependence of the Debye-Waller factor, the emission lineshape is modeled for temperatures from 78~K to 123~K (See figure S2 in supplementary information). The emission lineshape above this temperature could not be obtained with the theoretical model given above due to limitation of the two-phonon processes. Figure~\ref{fig:Fig3} (b) shows the ZPL fractions obtained from each spectrum, which tends to decrease linearly within the measured temperature range. From the extrapolation of the linear behavior,  the Debye-Waller factor at $T~=~0$~K is estimated as 0.59, which is much less than previously reported values for defects in hBN \cite{Jungwirth2016,Exarhos2017}. 

In conclusion, temperature-dependent spectral properties of a single defect in hBN are investigated both experimentally and theoretically. It is shown that the linear and quadratic electron-phonon coupling with acoustic phonons via the deformation potential and the piezoelectric coupling play significant role on the linewidth,  lineshift, and lineshape of the ZPL emission. We found that the linewidth of ZPL varies with  $T+T^5$ while its energy shifts with $T+T^3$.   In particular, the temperature-dependent lineshape analysis of the ZPL has yielded the Debye-Waller factor as low as 0.59 at 0~K. Finally, in addition to the ZPL, optical phonon sidebands assisted by the emission (Stokes) and absorption (anti-Stokes) of the low-energy Raman active $E_{2g}$ phonon mode is observed for the first time from a low temperature PL spectrum of a single defect in hBN.

\bibliographystyle{apsrev4-1}
\bibliography{Library_SA}

\begin{thebibliography}{35}%
\makeatletter
\providecommand \@ifxundefined [1]{%
 \@ifx{#1\undefined}
}%
\providecommand \@ifnum [1]{%
 \ifnum #1\expandafter \@firstoftwo
 \else \expandafter \@secondoftwo
 \fi
}%
\providecommand \@ifx [1]{%
 \ifx #1\expandafter \@firstoftwo
 \else \expandafter \@secondoftwo
 \fi
}%
\providecommand \natexlab [1]{#1}%
\providecommand \enquote  [1]{``#1''}%
\providecommand \bibnamefont  [1]{#1}%
\providecommand \bibfnamefont [1]{#1}%
\providecommand \citenamefont [1]{#1}%
\providecommand \href@noop [0]{\@secondoftwo}%
\providecommand \href [0]{\begingroup \@sanitize@url \@href}%
\providecommand \@href[1]{\@@startlink{#1}\@@href}%
\providecommand \@@href[1]{\endgroup#1\@@endlink}%
\providecommand \@sanitize@url [0]{\catcode `\\12\catcode `\$12\catcode
  `\&12\catcode `\#12\catcode `\^12\catcode `\_12\catcode `\%12\relax}%
\providecommand \@@startlink[1]{}%
\providecommand \@@endlink[0]{}%
\providecommand \url  [0]{\begingroup\@sanitize@url \@url }%
\providecommand \@url [1]{\endgroup\@href {#1}{\urlprefix }}%
\providecommand \urlprefix  [0]{URL }%
\providecommand \Eprint [0]{\href }%
\providecommand \doibase [0]{https://doi.org/}%
\providecommand \selectlanguage [0]{\@gobble}%
\providecommand \bibinfo  [0]{\@secondoftwo}%
\providecommand \bibfield  [0]{\@secondoftwo}%
\providecommand \translation [1]{[#1]}%
\providecommand \BibitemOpen [0]{}%
\providecommand \bibitemStop [0]{}%
\providecommand \bibitemNoStop [0]{.\EOS\space}%
\providecommand \EOS [0]{\spacefactor3000\relax}%
\providecommand \BibitemShut  [1]{\csname bibitem#1\endcsname}%
\let\auto@bib@innerbib\@empty
\bibitem [{\citenamefont {Aharonovich}\ \emph {et~al.}(2016)\citenamefont
  {Aharonovich}, \citenamefont {Englund},\ and\ \citenamefont
  {Toth}}]{Aharonovich2016}%
  \BibitemOpen
  \bibfield  {author} {\bibinfo {author} {\bibfnamefont {I.}~\bibnamefont
  {Aharonovich}}, \bibinfo {author} {\bibfnamefont {D.}~\bibnamefont
  {Englund}},\ and\ \bibinfo {author} {\bibfnamefont {M.}~\bibnamefont
  {Toth}},\ }\href@noop {} {\bibfield  {journal} {\bibinfo  {journal} {Nature
  Photonics}\ }\textbf {\bibinfo {volume} {10}},\ \bibinfo {pages} {631}
  (\bibinfo {year} {2016})}\BibitemShut {NoStop}%
\bibitem [{\citenamefont {He}\ \emph {et~al.}(2015)\citenamefont {He},
  \citenamefont {Clark}, \citenamefont {Schaibley}, \citenamefont {He},
  \citenamefont {Chen}, \citenamefont {Wei}, \citenamefont {Ding},
  \citenamefont {Zhang}, \citenamefont {Yao}, \citenamefont {Xu}, \citenamefont
  {Lu},\ and\ \citenamefont {Pan}}]{He2015}%
  \BibitemOpen
  \bibfield  {author} {\bibinfo {author} {\bibfnamefont {Y.-M.}\ \bibnamefont
  {He}}, \bibinfo {author} {\bibfnamefont {G.}~\bibnamefont {Clark}}, \bibinfo
  {author} {\bibfnamefont {J.~R.}\ \bibnamefont {Schaibley}}, \bibinfo {author}
  {\bibfnamefont {Y.}~\bibnamefont {He}}, \bibinfo {author} {\bibfnamefont
  {M.-C.}\ \bibnamefont {Chen}}, \bibinfo {author} {\bibfnamefont {Y.-J.}\
  \bibnamefont {Wei}}, \bibinfo {author} {\bibfnamefont {X.}~\bibnamefont
  {Ding}}, \bibinfo {author} {\bibfnamefont {Q.}~\bibnamefont {Zhang}},
  \bibinfo {author} {\bibfnamefont {W.}~\bibnamefont {Yao}}, \bibinfo {author}
  {\bibfnamefont {X.}~\bibnamefont {Xu}}, \bibinfo {author} {\bibfnamefont
  {C.-Y.}\ \bibnamefont {Lu}},\ and\ \bibinfo {author} {\bibfnamefont {J.-W.}\
  \bibnamefont {Pan}},\ }\href@noop {} {\bibfield  {journal} {\bibinfo
  {journal} {Nature Nanotechnology}\ }\textbf {\bibinfo {volume} {10}},\
  \bibinfo {pages} {497} (\bibinfo {year} {2015})}\BibitemShut {NoStop}%
\bibitem [{\citenamefont {Srivastava}\ \emph {et~al.}(2015)\citenamefont
  {Srivastava}, \citenamefont {Sidler}, \citenamefont {Allain}, \citenamefont
  {Lembke}, \citenamefont {Kis},\ and\ \citenamefont
  {Imamoğlu}}]{Srivastava2015a}%
  \BibitemOpen
  \bibfield  {author} {\bibinfo {author} {\bibfnamefont {A.}~\bibnamefont
  {Srivastava}}, \bibinfo {author} {\bibfnamefont {M.}~\bibnamefont {Sidler}},
  \bibinfo {author} {\bibfnamefont {A.~V.}\ \bibnamefont {Allain}}, \bibinfo
  {author} {\bibfnamefont {D.~S.}\ \bibnamefont {Lembke}}, \bibinfo {author}
  {\bibfnamefont {A.}~\bibnamefont {Kis}},\ and\ \bibinfo {author}
  {\bibfnamefont {A.}~\bibnamefont {Imamoğlu}},\ }\href@noop {} {\bibfield
  {journal} {\bibinfo  {journal} {Nature Nanotechnology}\ }\textbf {\bibinfo
  {volume} {10}},\ \bibinfo {pages} {491} (\bibinfo {year} {2015})}\BibitemShut
  {NoStop}%
\bibitem [{\citenamefont {Chakraborty}\ \emph {et~al.}(2015)\citenamefont
  {Chakraborty}, \citenamefont {Kinnischtzke}, \citenamefont {Goodfellow},
  \citenamefont {Beams},\ and\ \citenamefont {Vamivakas}}]{Chakraborty2015}%
  \BibitemOpen
  \bibfield  {author} {\bibinfo {author} {\bibfnamefont {C.}~\bibnamefont
  {Chakraborty}}, \bibinfo {author} {\bibfnamefont {L.}~\bibnamefont
  {Kinnischtzke}}, \bibinfo {author} {\bibfnamefont {K.~M.}\ \bibnamefont
  {Goodfellow}}, \bibinfo {author} {\bibfnamefont {R.}~\bibnamefont {Beams}},\
  and\ \bibinfo {author} {\bibfnamefont {a.~N.}\ \bibnamefont {Vamivakas}},\
  }\href@noop {} {\bibfield  {journal} {\bibinfo  {journal} {Nature
  Nanotechnology}\ }\textbf {\bibinfo {volume} {10}},\ \bibinfo {pages} {507}
  (\bibinfo {year} {2015})}\BibitemShut {NoStop}%
\bibitem [{\citenamefont {Tran}\ \emph {et~al.}(2015)\citenamefont {Tran},
  \citenamefont {Bray}, \citenamefont {Ford}, \citenamefont {Toth},\ and\
  \citenamefont {Aharonovich}}]{Tran2015}%
  \BibitemOpen
  \bibfield  {author} {\bibinfo {author} {\bibfnamefont {T.~T.}\ \bibnamefont
  {Tran}}, \bibinfo {author} {\bibfnamefont {K.}~\bibnamefont {Bray}}, \bibinfo
  {author} {\bibfnamefont {M.~J.}\ \bibnamefont {Ford}}, \bibinfo {author}
  {\bibfnamefont {M.}~\bibnamefont {Toth}},\ and\ \bibinfo {author}
  {\bibfnamefont {I.}~\bibnamefont {Aharonovich}},\ }\href
  {https://doi.org/10.1038/nnano.2015.242} {\bibfield  {journal} {\bibinfo
  {journal} {Nature Nanotechnology}\ }\textbf {\bibinfo {volume} {11}},\
  \bibinfo {pages} {37} (\bibinfo {year} {2015})}\BibitemShut {NoStop}%
\bibitem [{\citenamefont {Kianinia}\ \emph {et~al.}(2017)\citenamefont
  {Kianinia}, \citenamefont {Regan}, \citenamefont {Tawfik}, \citenamefont
  {Tran}, \citenamefont {Ford}, \citenamefont {Aharonovich},\ and\
  \citenamefont {Toth}}]{Kianinia2017}%
  \BibitemOpen
  \bibfield  {author} {\bibinfo {author} {\bibfnamefont {M.}~\bibnamefont
  {Kianinia}}, \bibinfo {author} {\bibfnamefont {B.}~\bibnamefont {Regan}},
  \bibinfo {author} {\bibfnamefont {S.~A.}\ \bibnamefont {Tawfik}}, \bibinfo
  {author} {\bibfnamefont {T.~T.}\ \bibnamefont {Tran}}, \bibinfo {author}
  {\bibfnamefont {M.~J.}\ \bibnamefont {Ford}}, \bibinfo {author}
  {\bibfnamefont {I.}~\bibnamefont {Aharonovich}},\ and\ \bibinfo {author}
  {\bibfnamefont {M.}~\bibnamefont {Toth}},\ }\href
  {https://doi.org/10.1021/acsphotonics.7b00086} {\bibfield  {journal}
  {\bibinfo  {journal} {ACS Photonics}\ }\textbf {\bibinfo {volume} {4}},\
  \bibinfo {pages} {768} (\bibinfo {year} {2017})}\BibitemShut {NoStop}%
\bibitem [{\citenamefont {Jungwirth}\ \emph {et~al.}(2016)\citenamefont
  {Jungwirth}, \citenamefont {Calderon}, \citenamefont {Ji}, \citenamefont
  {Spencer}, \citenamefont {Flatt{\'{e}}},\ and\ \citenamefont
  {Fuchs}}]{Jungwirth2016}%
  \BibitemOpen
  \bibfield  {author} {\bibinfo {author} {\bibfnamefont {N.~R.}\ \bibnamefont
  {Jungwirth}}, \bibinfo {author} {\bibfnamefont {B.}~\bibnamefont {Calderon}},
  \bibinfo {author} {\bibfnamefont {Y.}~\bibnamefont {Ji}}, \bibinfo {author}
  {\bibfnamefont {M.~G.}\ \bibnamefont {Spencer}}, \bibinfo {author}
  {\bibfnamefont {M.~E.}\ \bibnamefont {Flatt{\'{e}}}},\ and\ \bibinfo {author}
  {\bibfnamefont {G.~D.}\ \bibnamefont {Fuchs}},\ }\href@noop {} {\bibfield
  {journal} {\bibinfo  {journal} {Nano Letters}\ }\textbf {\bibinfo {volume}
  {16}},\ \bibinfo {pages} {6052} (\bibinfo {year} {2016})}\BibitemShut
  {NoStop}%
\bibitem [{\citenamefont {Kim}\ \emph {et~al.}(2018)\citenamefont {Kim},
  \citenamefont {Fr{\"{o}}ch}, \citenamefont {Christian}, \citenamefont
  {Straw}, \citenamefont {Bishop}, \citenamefont {Totonjian}, \citenamefont
  {Watanabe}, \citenamefont {Taniguchi}, \citenamefont {Toth},\ and\
  \citenamefont {Aharonovich}}]{Kim2018b}%
  \BibitemOpen
  \bibfield  {author} {\bibinfo {author} {\bibfnamefont {S.}~\bibnamefont
  {Kim}}, \bibinfo {author} {\bibfnamefont {J.~E.}\ \bibnamefont
  {Fr{\"{o}}ch}}, \bibinfo {author} {\bibfnamefont {J.}~\bibnamefont
  {Christian}}, \bibinfo {author} {\bibfnamefont {M.}~\bibnamefont {Straw}},
  \bibinfo {author} {\bibfnamefont {J.}~\bibnamefont {Bishop}}, \bibinfo
  {author} {\bibfnamefont {D.}~\bibnamefont {Totonjian}}, \bibinfo {author}
  {\bibfnamefont {K.}~\bibnamefont {Watanabe}}, \bibinfo {author}
  {\bibfnamefont {T.}~\bibnamefont {Taniguchi}}, \bibinfo {author}
  {\bibfnamefont {M.}~\bibnamefont {Toth}},\ and\ \bibinfo {author}
  {\bibfnamefont {I.}~\bibnamefont {Aharonovich}},\ }\href
  {https://doi.org/10.1038/s41467-018-05117-4} {\bibfield  {journal} {\bibinfo
  {journal} {Nature Communications}\ }\textbf {\bibinfo {volume} {9}},\
  \bibinfo {pages} {2623} (\bibinfo {year} {2018})}\BibitemShut {NoStop}%
\bibitem [{\citenamefont {Vuong}\ \emph {et~al.}(2016)\citenamefont {Vuong},
  \citenamefont {Cassabois}, \citenamefont {Valvin}, \citenamefont {Ouerghi},
  \citenamefont {Chassagneux}, \citenamefont {Voisin},\ and\ \citenamefont
  {Gil}}]{Vuong2016}%
  \BibitemOpen
  \bibfield  {author} {\bibinfo {author} {\bibfnamefont {T.~Q.~P.}\
  \bibnamefont {Vuong}}, \bibinfo {author} {\bibfnamefont {G.}~\bibnamefont
  {Cassabois}}, \bibinfo {author} {\bibfnamefont {P.}~\bibnamefont {Valvin}},
  \bibinfo {author} {\bibfnamefont {A.}~\bibnamefont {Ouerghi}}, \bibinfo
  {author} {\bibfnamefont {Y.}~\bibnamefont {Chassagneux}}, \bibinfo {author}
  {\bibfnamefont {C.}~\bibnamefont {Voisin}},\ and\ \bibinfo {author}
  {\bibfnamefont {B.}~\bibnamefont {Gil}},\ }\href@noop {} {\bibfield
  {journal} {\bibinfo  {journal} {Physical Review Letters}\ }\textbf {\bibinfo
  {volume} {117}},\ \bibinfo {pages} {097402} (\bibinfo {year}
  {2016})}\BibitemShut {NoStop}%
\bibitem [{\citenamefont {Jungwirth}\ and\ \citenamefont
  {Fuchs}(2017)}]{Jungwirth2017}%
  \BibitemOpen
  \bibfield  {author} {\bibinfo {author} {\bibfnamefont {N.~R.}\ \bibnamefont
  {Jungwirth}}\ and\ \bibinfo {author} {\bibfnamefont {G.~D.}\ \bibnamefont
  {Fuchs}},\ }\href@noop {} {\bibfield  {journal} {\bibinfo  {journal}
  {Physical Review Letters}\ }\textbf {\bibinfo {volume} {119}},\ \bibinfo
  {pages} {057401} (\bibinfo {year} {2017})}\BibitemShut {NoStop}%
\bibitem [{\citenamefont {Exarhos}\ \emph {et~al.}(2017)\citenamefont
  {Exarhos}, \citenamefont {Hopper}, \citenamefont {Grote}, \citenamefont
  {Alkauskas},\ and\ \citenamefont {Bassett}}]{Exarhos2017}%
  \BibitemOpen
  \bibfield  {author} {\bibinfo {author} {\bibfnamefont {A.~L.}\ \bibnamefont
  {Exarhos}}, \bibinfo {author} {\bibfnamefont {D.~A.}\ \bibnamefont {Hopper}},
  \bibinfo {author} {\bibfnamefont {R.~R.}\ \bibnamefont {Grote}}, \bibinfo
  {author} {\bibfnamefont {A.}~\bibnamefont {Alkauskas}},\ and\ \bibinfo
  {author} {\bibfnamefont {L.~C.}\ \bibnamefont {Bassett}},\ }\href
  {https://doi.org/10.1021/acsnano.7b00665} {\bibfield  {journal} {\bibinfo
  {journal} {ACS Nano}\ }\textbf {\bibinfo {volume} {11}},\ \bibinfo {pages}
  {3328} (\bibinfo {year} {2017})}\BibitemShut {NoStop}%
\bibitem [{\citenamefont {Geick}\ \emph {et~al.}(1966)\citenamefont {Geick},
  \citenamefont {Perry},\ and\ \citenamefont {Rupprecht}}]{Geick1966}%
  \BibitemOpen
  \bibfield  {author} {\bibinfo {author} {\bibfnamefont {R.}~\bibnamefont
  {Geick}}, \bibinfo {author} {\bibfnamefont {C.~H.}\ \bibnamefont {Perry}},\
  and\ \bibinfo {author} {\bibfnamefont {G.}~\bibnamefont {Rupprecht}},\ }\href
  {https://doi.org/10.1103/PhysRev.146.543} {\bibfield  {journal} {\bibinfo
  {journal} {Physical Review}\ }\textbf {\bibinfo {volume} {146}},\ \bibinfo
  {pages} {543} (\bibinfo {year} {1966})}\BibitemShut {NoStop}%
\bibitem [{\citenamefont {Kuzuba}\ \emph {et~al.}(1978)\citenamefont {Kuzuba},
  \citenamefont {Era}, \citenamefont {Ishii},\ and\ \citenamefont
  {Sato}}]{Kuzuba1978}%
  \BibitemOpen
  \bibfield  {author} {\bibinfo {author} {\bibfnamefont {T.}~\bibnamefont
  {Kuzuba}}, \bibinfo {author} {\bibfnamefont {K.}~\bibnamefont {Era}},
  \bibinfo {author} {\bibfnamefont {T.}~\bibnamefont {Ishii}},\ and\ \bibinfo
  {author} {\bibfnamefont {T.}~\bibnamefont {Sato}},\ }\href
  {https://doi.org/10.1016/0038-1098(78)90288-0} {\bibfield  {journal}
  {\bibinfo  {journal} {Solid State Communications}\ }\textbf {\bibinfo
  {volume} {25}},\ \bibinfo {pages} {863} (\bibinfo {year} {1978})}\BibitemShut
  {NoStop}%
\bibitem [{\citenamefont {Nemanich}\ \emph {et~al.}(1981)\citenamefont
  {Nemanich}, \citenamefont {Solin},\ and\ \citenamefont
  {Martin}}]{Nemanich1981}%
  \BibitemOpen
  \bibfield  {author} {\bibinfo {author} {\bibfnamefont {R.~J.}\ \bibnamefont
  {Nemanich}}, \bibinfo {author} {\bibfnamefont {S.~A.}\ \bibnamefont
  {Solin}},\ and\ \bibinfo {author} {\bibfnamefont {R.~M.}\ \bibnamefont
  {Martin}},\ }\href {https://doi.org/10.1103/PhysRevB.23.6348} {\bibfield
  {journal} {\bibinfo  {journal} {Physical Review B}\ }\textbf {\bibinfo
  {volume} {23}},\ \bibinfo {pages} {6348} (\bibinfo {year}
  {1981})}\BibitemShut {NoStop}%
\bibitem [{\citenamefont {Cusc{\'{o}}}\ \emph {et~al.}(2016)\citenamefont
  {Cusc{\'{o}}}, \citenamefont {Gil}, \citenamefont {Cassabois},\ and\
  \citenamefont {Art{\'{u}}s}}]{Cusco2016}%
  \BibitemOpen
  \bibfield  {author} {\bibinfo {author} {\bibfnamefont {R.}~\bibnamefont
  {Cusc{\'{o}}}}, \bibinfo {author} {\bibfnamefont {B.}~\bibnamefont {Gil}},
  \bibinfo {author} {\bibfnamefont {G.}~\bibnamefont {Cassabois}},\ and\
  \bibinfo {author} {\bibfnamefont {L.}~\bibnamefont {Art{\'{u}}s}},\
  }\href@noop {} {\bibfield  {journal} {\bibinfo  {journal} {Physical Review
  B}\ }\textbf {\bibinfo {volume} {94}},\ \bibinfo {pages} {155435} (\bibinfo
  {year} {2016})}\BibitemShut {NoStop}%
\bibitem [{\citenamefont {Mart{\'{i}}nez}\ \emph {et~al.}(2016)\citenamefont
  {Mart{\'{i}}nez}, \citenamefont {Pelini}, \citenamefont {Waselowski},
  \citenamefont {Maze}, \citenamefont {Gil}, \citenamefont {Cassabois},\ and\
  \citenamefont {Jacques}}]{Martinez2016}%
  \BibitemOpen
  \bibfield  {author} {\bibinfo {author} {\bibfnamefont {L.~J.}\ \bibnamefont
  {Mart{\'{i}}nez}}, \bibinfo {author} {\bibfnamefont {T.}~\bibnamefont
  {Pelini}}, \bibinfo {author} {\bibfnamefont {V.}~\bibnamefont {Waselowski}},
  \bibinfo {author} {\bibfnamefont {J.~R.}\ \bibnamefont {Maze}}, \bibinfo
  {author} {\bibfnamefont {B.}~\bibnamefont {Gil}}, \bibinfo {author}
  {\bibfnamefont {G.}~\bibnamefont {Cassabois}},\ and\ \bibinfo {author}
  {\bibfnamefont {V.}~\bibnamefont {Jacques}},\ }\href
  {https://doi.org/10.1103/PhysRevB.94.121405} {\bibfield  {journal} {\bibinfo
  {journal} {Physical Review B}\ }\textbf {\bibinfo {volume} {94}},\ \bibinfo
  {pages} {121405} (\bibinfo {year} {2016})}\BibitemShut {NoStop}%
\bibitem [{\citenamefont {Vuong}\ \emph {et~al.}(2017)\citenamefont {Vuong},
  \citenamefont {Cassabois}, \citenamefont {Valvin}, \citenamefont {Jacques},
  \citenamefont {Cusc{\'{o}}}, \citenamefont {Art{\'{u}}s},\ and\ \citenamefont
  {Gil}}]{Vuong2017a}%
  \BibitemOpen
  \bibfield  {author} {\bibinfo {author} {\bibfnamefont {T.~Q.~P.}\
  \bibnamefont {Vuong}}, \bibinfo {author} {\bibfnamefont {G.}~\bibnamefont
  {Cassabois}}, \bibinfo {author} {\bibfnamefont {P.}~\bibnamefont {Valvin}},
  \bibinfo {author} {\bibfnamefont {V.}~\bibnamefont {Jacques}}, \bibinfo
  {author} {\bibfnamefont {R.}~\bibnamefont {Cusc{\'{o}}}}, \bibinfo {author}
  {\bibfnamefont {L.}~\bibnamefont {Art{\'{u}}s}},\ and\ \bibinfo {author}
  {\bibfnamefont {B.}~\bibnamefont {Gil}},\ }\href@noop {} {\bibfield
  {journal} {\bibinfo  {journal} {Physical Review B}\ }\textbf {\bibinfo
  {volume} {95}},\ \bibinfo {pages} {045207} (\bibinfo {year}
  {2017})}\BibitemShut {NoStop}%
\bibitem [{\citenamefont {Stenger}\ \emph {et~al.}(2017)\citenamefont
  {Stenger}, \citenamefont {Schue}, \citenamefont {Boukhicha}, \citenamefont
  {Berini}, \citenamefont {Pla?ais}, \citenamefont {Loiseau},\ and\
  \citenamefont {Barjon}}]{Stenger2017}%
  \BibitemOpen
  \bibfield  {author} {\bibinfo {author} {\bibfnamefont {I.}~\bibnamefont
  {Stenger}}, \bibinfo {author} {\bibfnamefont {L.}~\bibnamefont {Schue}},
  \bibinfo {author} {\bibfnamefont {M.}~\bibnamefont {Boukhicha}}, \bibinfo
  {author} {\bibfnamefont {B.}~\bibnamefont {Berini}}, \bibinfo {author}
  {\bibfnamefont {B.}~\bibnamefont {Pla?ais}}, \bibinfo {author} {\bibfnamefont
  {A.}~\bibnamefont {Loiseau}},\ and\ \bibinfo {author} {\bibfnamefont
  {J.}~\bibnamefont {Barjon}},\ }\href
  {https://doi.org/10.1088/2053-1583/aa77d4} {\bibfield  {journal} {\bibinfo
  {journal} {2D Materials}\ }\textbf {\bibinfo {volume} {4}},\ \bibinfo {pages}
  {031003} (\bibinfo {year} {2017})}\BibitemShut {NoStop}%
\bibitem [{\citenamefont {Norambuena}\ \emph {et~al.}(2016)\citenamefont
  {Norambuena}, \citenamefont {Reyes}, \citenamefont {Mej{\'{i}}a-Lop{\'{e}}z},
  \citenamefont {Gali},\ and\ \citenamefont {Maze}}]{Norambuena2016}%
  \BibitemOpen
  \bibfield  {author} {\bibinfo {author} {\bibfnamefont {A.}~\bibnamefont
  {Norambuena}}, \bibinfo {author} {\bibfnamefont {S.~A.}\ \bibnamefont
  {Reyes}}, \bibinfo {author} {\bibfnamefont {J.}~\bibnamefont
  {Mej{\'{i}}a-Lop{\'{e}}z}}, \bibinfo {author} {\bibfnamefont
  {A.}~\bibnamefont {Gali}},\ and\ \bibinfo {author} {\bibfnamefont {J.~R.}\
  \bibnamefont {Maze}},\ }\href@noop {} {\bibfield  {journal} {\bibinfo
  {journal} {Physical Review B}\ }\textbf {\bibinfo {volume} {94}},\ \bibinfo
  {pages} {134305} (\bibinfo {year} {2016})}\BibitemShut {NoStop}%
\bibitem [{\citenamefont {Paszkowicz}\ \emph {et~al.}(2002)\citenamefont
  {Paszkowicz}, \citenamefont {Pelka}, \citenamefont {Knapp}, \citenamefont
  {Szyszko},\ and\ \citenamefont {Podsiadlo}}]{Paszkowicz2002}%
  \BibitemOpen
  \bibfield  {author} {\bibinfo {author} {\bibfnamefont {W.}~\bibnamefont
  {Paszkowicz}}, \bibinfo {author} {\bibfnamefont {J.}~\bibnamefont {Pelka}},
  \bibinfo {author} {\bibfnamefont {M.}~\bibnamefont {Knapp}}, \bibinfo
  {author} {\bibfnamefont {T.}~\bibnamefont {Szyszko}},\ and\ \bibinfo {author}
  {\bibfnamefont {S.}~\bibnamefont {Podsiadlo}},\ }\href@noop {} {\bibfield
  {journal} {\bibinfo  {journal} {Applied Physics A: Materials Science {\&}
  Processing}\ }\textbf {\bibinfo {volume} {75}},\ \bibinfo {pages} {431}
  (\bibinfo {year} {2002})}\BibitemShut {NoStop}%
\bibitem [{\citenamefont {M{\"{u}}ller}\ \emph {et~al.}(2012)\citenamefont
  {M{\"{u}}ller}, \citenamefont {Aharonovich}, \citenamefont {Wang},
  \citenamefont {Yuan}, \citenamefont {Castelletto}, \citenamefont {Prawer},\
  and\ \citenamefont {Atat{\"{u}}re}}]{Muller2012}%
  \BibitemOpen
  \bibfield  {author} {\bibinfo {author} {\bibfnamefont {T.}~\bibnamefont
  {M{\"{u}}ller}}, \bibinfo {author} {\bibfnamefont {I.}~\bibnamefont
  {Aharonovich}}, \bibinfo {author} {\bibfnamefont {Z.}~\bibnamefont {Wang}},
  \bibinfo {author} {\bibfnamefont {X.}~\bibnamefont {Yuan}}, \bibinfo {author}
  {\bibfnamefont {S.}~\bibnamefont {Castelletto}}, \bibinfo {author}
  {\bibfnamefont {S.}~\bibnamefont {Prawer}},\ and\ \bibinfo {author}
  {\bibfnamefont {M.}~\bibnamefont {Atat{\"{u}}re}},\ }\href
  {https://doi.org/10.1103/PhysRevB.86.195210} {\bibfield  {journal} {\bibinfo
  {journal} {Physical Review B}\ }\textbf {\bibinfo {volume} {86}},\ \bibinfo
  {pages} {195210} (\bibinfo {year} {2012})}\BibitemShut {NoStop}%
\bibitem [{\citenamefont {Neu}\ \emph {et~al.}(2013)\citenamefont {Neu},
  \citenamefont {Hepp}, \citenamefont {Hauschild}, \citenamefont {Gsell},
  \citenamefont {Fischer}, \citenamefont {Sternschulte}, \citenamefont
  {Steinm{\"{u}}ller-Nethl}, \citenamefont {Schreck},\ and\ \citenamefont
  {Becher}}]{Neu2013}%
  \BibitemOpen
  \bibfield  {author} {\bibinfo {author} {\bibfnamefont {E.}~\bibnamefont
  {Neu}}, \bibinfo {author} {\bibfnamefont {C.}~\bibnamefont {Hepp}}, \bibinfo
  {author} {\bibfnamefont {M.}~\bibnamefont {Hauschild}}, \bibinfo {author}
  {\bibfnamefont {S.}~\bibnamefont {Gsell}}, \bibinfo {author} {\bibfnamefont
  {M.}~\bibnamefont {Fischer}}, \bibinfo {author} {\bibfnamefont
  {H.}~\bibnamefont {Sternschulte}}, \bibinfo {author} {\bibfnamefont
  {D.}~\bibnamefont {Steinm{\"{u}}ller-Nethl}}, \bibinfo {author}
  {\bibfnamefont {M.}~\bibnamefont {Schreck}},\ and\ \bibinfo {author}
  {\bibfnamefont {C.}~\bibnamefont {Becher}},\ }\href@noop {} {\bibfield
  {journal} {\bibinfo  {journal} {New Journal of Physics}\ }\textbf {\bibinfo
  {volume} {15}},\ \bibinfo {pages} {043005} (\bibinfo {year}
  {2013})}\BibitemShut {NoStop}%
\bibitem [{\citenamefont {Dietrich}\ \emph {et~al.}(2018)\citenamefont
  {Dietrich}, \citenamefont {B{\"{u}}rk}, \citenamefont {Steiger},
  \citenamefont {Antoniuk}, \citenamefont {Tran}, \citenamefont {Nguyen},
  \citenamefont {Aharonovich}, \citenamefont {Jelezko},\ and\ \citenamefont
  {Kubanek}}]{Dietrich2018}%
  \BibitemOpen
  \bibfield  {author} {\bibinfo {author} {\bibfnamefont {A.}~\bibnamefont
  {Dietrich}}, \bibinfo {author} {\bibfnamefont {M.}~\bibnamefont
  {B{\"{u}}rk}}, \bibinfo {author} {\bibfnamefont {E.~S.}\ \bibnamefont
  {Steiger}}, \bibinfo {author} {\bibfnamefont {L.}~\bibnamefont {Antoniuk}},
  \bibinfo {author} {\bibfnamefont {T.~T.}\ \bibnamefont {Tran}}, \bibinfo
  {author} {\bibfnamefont {M.}~\bibnamefont {Nguyen}}, \bibinfo {author}
  {\bibfnamefont {I.}~\bibnamefont {Aharonovich}}, \bibinfo {author}
  {\bibfnamefont {F.}~\bibnamefont {Jelezko}},\ and\ \bibinfo {author}
  {\bibfnamefont {A.}~\bibnamefont {Kubanek}},\ }\href
  {https://doi.org/10.1103/PhysRevB.98.081414} {\bibfield  {journal} {\bibinfo
  {journal} {Physical Review B}\ }\textbf {\bibinfo {volume} {98}},\ \bibinfo
  {pages} {081414} (\bibinfo {year} {2018})}\BibitemShut {NoStop}%
\bibitem [{\citenamefont {Wolters}\ \emph {et~al.}(2013)\citenamefont
  {Wolters}, \citenamefont {Sadzak}, \citenamefont {Schell}, \citenamefont
  {Schr{\"{o}}der},\ and\ \citenamefont {Benson}}]{Wolters2013}%
  \BibitemOpen
  \bibfield  {author} {\bibinfo {author} {\bibfnamefont {J.}~\bibnamefont
  {Wolters}}, \bibinfo {author} {\bibfnamefont {N.}~\bibnamefont {Sadzak}},
  \bibinfo {author} {\bibfnamefont {A.~W.}\ \bibnamefont {Schell}}, \bibinfo
  {author} {\bibfnamefont {T.}~\bibnamefont {Schr{\"{o}}der}},\ and\ \bibinfo
  {author} {\bibfnamefont {O.}~\bibnamefont {Benson}},\ }\href
  {https://doi.org/10.1103/PhysRevLett.110.027401} {\bibfield  {journal}
  {\bibinfo  {journal} {Physical Review Letters}\ }\textbf {\bibinfo {volume}
  {110}},\ \bibinfo {pages} {027401} (\bibinfo {year} {2013})}\BibitemShut
  {NoStop}%
\bibitem [{\citenamefont {Fu}\ \emph {et~al.}(2009)\citenamefont {Fu},
  \citenamefont {Santori}, \citenamefont {Barclay}, \citenamefont {Rogers},
  \citenamefont {Manson},\ and\ \citenamefont {Beausoleil}}]{Fu2009}%
  \BibitemOpen
  \bibfield  {author} {\bibinfo {author} {\bibfnamefont {K.-M.~C.}\
  \bibnamefont {Fu}}, \bibinfo {author} {\bibfnamefont {C.}~\bibnamefont
  {Santori}}, \bibinfo {author} {\bibfnamefont {P.~E.}\ \bibnamefont
  {Barclay}}, \bibinfo {author} {\bibfnamefont {L.~J.}\ \bibnamefont {Rogers}},
  \bibinfo {author} {\bibfnamefont {N.~B.}\ \bibnamefont {Manson}},\ and\
  \bibinfo {author} {\bibfnamefont {R.~G.}\ \bibnamefont {Beausoleil}},\
  }\href@noop {} {\bibfield  {journal} {\bibinfo  {journal} {Physical Review
  Letters}\ }\textbf {\bibinfo {volume} {103}},\ \bibinfo {pages} {256404}
  (\bibinfo {year} {2009})}\BibitemShut {NoStop}%
\bibitem [{\citenamefont {Favero}\ \emph {et~al.}(2003)\citenamefont {Favero},
  \citenamefont {Cassabois}, \citenamefont {Ferreira}, \citenamefont {Darson},
  \citenamefont {Voisin}, \citenamefont {Tignon}, \citenamefont {Delalande},
  \citenamefont {Bastard}, \citenamefont {Roussignol},\ and\ \citenamefont
  {G{\'{e}}rard}}]{Favero2003}%
  \BibitemOpen
  \bibfield  {author} {\bibinfo {author} {\bibfnamefont {I.}~\bibnamefont
  {Favero}}, \bibinfo {author} {\bibfnamefont {G.}~\bibnamefont {Cassabois}},
  \bibinfo {author} {\bibfnamefont {R.}~\bibnamefont {Ferreira}}, \bibinfo
  {author} {\bibfnamefont {D.}~\bibnamefont {Darson}}, \bibinfo {author}
  {\bibfnamefont {C.}~\bibnamefont {Voisin}}, \bibinfo {author} {\bibfnamefont
  {J.}~\bibnamefont {Tignon}}, \bibinfo {author} {\bibfnamefont
  {C.}~\bibnamefont {Delalande}}, \bibinfo {author} {\bibfnamefont
  {G.}~\bibnamefont {Bastard}}, \bibinfo {author} {\bibfnamefont
  {P.}~\bibnamefont {Roussignol}},\ and\ \bibinfo {author} {\bibfnamefont
  {J.~M.}\ \bibnamefont {G{\'{e}}rard}},\ }\href@noop {} {\bibfield  {journal}
  {\bibinfo  {journal} {Physical Review B}\ }\textbf {\bibinfo {volume} {68}},\
  \bibinfo {pages} {233301} (\bibinfo {year} {2003})}\BibitemShut {NoStop}%
\bibitem [{\citenamefont {Besombes}\ \emph {et~al.}(2001)\citenamefont
  {Besombes}, \citenamefont {Kheng}, \citenamefont {Marsal},\ and\
  \citenamefont {Mariette}}]{Besombes2001}%
  \BibitemOpen
  \bibfield  {author} {\bibinfo {author} {\bibfnamefont {L.}~\bibnamefont
  {Besombes}}, \bibinfo {author} {\bibfnamefont {K.}~\bibnamefont {Kheng}},
  \bibinfo {author} {\bibfnamefont {L.}~\bibnamefont {Marsal}},\ and\ \bibinfo
  {author} {\bibfnamefont {H.}~\bibnamefont {Mariette}},\ }\href@noop {}
  {\bibfield  {journal} {\bibinfo  {journal} {Physical Review B}\ }\textbf
  {\bibinfo {volume} {63}},\ \bibinfo {pages} {155307} (\bibinfo {year}
  {2001})}\BibitemShut {NoStop}%
\bibitem [{\citenamefont {Hizhnyakov}\ \emph {et~al.}(2002)\citenamefont
  {Hizhnyakov}, \citenamefont {Kaasik},\ and\ \citenamefont
  {Sildos}}]{Hizhnyakov2002}%
  \BibitemOpen
  \bibfield  {author} {\bibinfo {author} {\bibfnamefont {V.}~\bibnamefont
  {Hizhnyakov}}, \bibinfo {author} {\bibfnamefont {H.}~\bibnamefont {Kaasik}},\
  and\ \bibinfo {author} {\bibfnamefont {I.}~\bibnamefont {Sildos}},\
  }\href@noop {} {\bibfield  {journal} {\bibinfo  {journal} {Physica Status
  Solidi (B) Basic Research}\ }\textbf {\bibinfo {volume} {234}},\ \bibinfo
  {pages} {644} (\bibinfo {year} {2002})}\BibitemShut {NoStop}%
\bibitem [{\citenamefont {Davies}(1974)}]{Davies1974}%
  \BibitemOpen
  \bibfield  {author} {\bibinfo {author} {\bibfnamefont {G.}~\bibnamefont
  {Davies}},\ }\href@noop {} {\bibfield  {journal} {\bibinfo  {journal}
  {Journal of Physics C: Solid State Physics}\ }\textbf {\bibinfo {volume}
  {7}},\ \bibinfo {pages} {3797} (\bibinfo {year} {1974})}\BibitemShut
  {NoStop}%
\bibitem [{\citenamefont {Stonehamm}(2001)}]{Stonehamm2001}%
  \BibitemOpen
  \bibfield  {author} {\bibinfo {author} {\bibfnamefont {A.~M.}\ \bibnamefont
  {Stonehamm}},\ }\href@noop {} {\emph {\bibinfo {title} {{Theory of Defects in
  Solids}}}}\ (\bibinfo {year} {2001})\BibitemShut {NoStop}%
\bibitem [{\citenamefont {Tohei}\ \emph {et~al.}(2006)\citenamefont {Tohei},
  \citenamefont {Kuwabara}, \citenamefont {Oba},\ and\ \citenamefont
  {Tanaka}}]{Tohei2006}%
  \BibitemOpen
  \bibfield  {author} {\bibinfo {author} {\bibfnamefont {T.}~\bibnamefont
  {Tohei}}, \bibinfo {author} {\bibfnamefont {A.}~\bibnamefont {Kuwabara}},
  \bibinfo {author} {\bibfnamefont {F.}~\bibnamefont {Oba}},\ and\ \bibinfo
  {author} {\bibfnamefont {I.}~\bibnamefont {Tanaka}},\ }\href@noop {}
  {\bibfield  {journal} {\bibinfo  {journal} {Physical Review B}\ }\textbf
  {\bibinfo {volume} {73}},\ \bibinfo {pages} {064304} (\bibinfo {year}
  {2006})}\BibitemShut {NoStop}%
\bibitem [{\citenamefont {Sontheimer}\ \emph {et~al.}(2017)\citenamefont
  {Sontheimer}, \citenamefont {Braun}, \citenamefont {Nikolay}, \citenamefont
  {Sadzak}, \citenamefont {Aharonovich},\ and\ \citenamefont
  {Benson}}]{Sontheimer2017}%
  \BibitemOpen
  \bibfield  {author} {\bibinfo {author} {\bibfnamefont {B.}~\bibnamefont
  {Sontheimer}}, \bibinfo {author} {\bibfnamefont {M.}~\bibnamefont {Braun}},
  \bibinfo {author} {\bibfnamefont {N.}~\bibnamefont {Nikolay}}, \bibinfo
  {author} {\bibfnamefont {N.}~\bibnamefont {Sadzak}}, \bibinfo {author}
  {\bibfnamefont {I.}~\bibnamefont {Aharonovich}},\ and\ \bibinfo {author}
  {\bibfnamefont {O.}~\bibnamefont {Benson}},\ }\href@noop {} {\bibfield
  {journal} {\bibinfo  {journal} {Physical Review B}\ }\textbf {\bibinfo
  {volume} {96}},\ \bibinfo {pages} {121202} (\bibinfo {year}
  {2017})}\BibitemShut {NoStop}%
\bibitem [{\citenamefont {Keil}(1965)}]{Keil1965}%
  \BibitemOpen
  \bibfield  {author} {\bibinfo {author} {\bibfnamefont {T.~H.}\ \bibnamefont
  {Keil}},\ }\href {https://doi.org/10.1103/PhysRev.140.A601} {\bibfield
  {journal} {\bibinfo  {journal} {Physical Review}\ }\textbf {\bibinfo {volume}
  {140}},\ \bibinfo {pages} {A601} (\bibinfo {year} {1965})}\BibitemShut
  {NoStop}%
\bibitem [{\citenamefont {Krummheuer}\ \emph {et~al.}(2002)\citenamefont
  {Krummheuer}, \citenamefont {Axt},\ and\ \citenamefont
  {Kuhn}}]{Krummheuer2002}%
  \BibitemOpen
  \bibfield  {author} {\bibinfo {author} {\bibfnamefont {B.}~\bibnamefont
  {Krummheuer}}, \bibinfo {author} {\bibfnamefont {V.~M.}\ \bibnamefont
  {Axt}},\ and\ \bibinfo {author} {\bibfnamefont {T.}~\bibnamefont {Kuhn}},\
  }\href {https://doi.org/10.1103/PhysRevB.65.195313} {\bibfield  {journal}
  {\bibinfo  {journal} {Physical Review B}\ }\textbf {\bibinfo {volume} {65}},\
  \bibinfo {pages} {195313} (\bibinfo {year} {2002})}\BibitemShut {NoStop}%
\bibitem [{\citenamefont {Alkauskas}\ \emph {et~al.}(2014)\citenamefont
  {Alkauskas}, \citenamefont {Buckley}, \citenamefont {Awschalom},\ and\
  \citenamefont {{Van de Walle}}}]{Alkauskas2014}%
  \BibitemOpen
  \bibfield  {author} {\bibinfo {author} {\bibfnamefont {A.}~\bibnamefont
  {Alkauskas}}, \bibinfo {author} {\bibfnamefont {B.~B.}\ \bibnamefont
  {Buckley}}, \bibinfo {author} {\bibfnamefont {D.~D.}\ \bibnamefont
  {Awschalom}},\ and\ \bibinfo {author} {\bibfnamefont {C.~G.}\ \bibnamefont
  {{Van de Walle}}},\ }\href@noop {} {\bibfield  {journal} {\bibinfo  {journal}
  {New Journal of Physics}\ }\textbf {\bibinfo {volume} {16}},\ \bibinfo
  {pages} {073026} (\bibinfo {year} {2014})}\BibitemShut {NoStop}%
\end{thebibliography}%


\begin{thebibliography}{1}
\expandafter\ifx\csname natexlab\endcsname\relax\def\natexlab#1{#1}\fi
\expandafter\ifx\csname bibnamefont\endcsname\relax
  \def\bibnamefont#1{#1}\fi
\expandafter\ifx\csname bibfnamefont\endcsname\relax
  \def\bibfnamefont#1{#1}\fi
\expandafter\ifx\csname citenamefont\endcsname\relax
  \def\citenamefont#1{#1}\fi
\expandafter\ifx\csname url\endcsname\relax
  \def\url#1{\texttt{#1}}\fi
\expandafter\ifx\csname urlprefix\endcsname\relax\def\urlprefix{URL }\fi
\providecommand{\bibinfo}[2]{#2}
\providecommand{\eprint}[2][]{\url{#2}}

\bibitem[{\citenamefont{Stoneham}(2001)}]{Stoneham2001}
\bibinfo{author}{\bibfnamefont{A.~M.} \bibnamefont{Stoneham}}, in
  \emph{\bibinfo{booktitle}{Theory of Defects in Solids}}
  (\bibinfo{year}{2001}).

\end{thebibliography}

\end{document}



\title{Supplemental Information for \\The effect of electron-phonon interactions on the spectral properties of single defects in hBN}

\author{Ozan Ar{\i}}\email{ozanari@iyte.edu.tr}
\affiliation{Department of Physics,
Izmir Institute of Technology, Urla, Turkey}
\author{Nahit Polat}
\affiliation{Department of Physics,
Izmir Institute of Technology, Urla, Turkey}
\author{Özgür Çakır}
\affiliation{Department of Physics,
Izmir Institute of Technology, Urla, Turkey}
\author{Serkan Ates}\email{serkanates@iyte.edu.tr}
\affiliation{Department of Physics,
Izmir Institute of Technology, Urla, Turkey}

\maketitle
\setcounter{figure}{0}

\renewcommand{\thefigure}{S\@arabic\c@figure}
\section{emission process from defects in hbn}
\begin{figure}[h]
\centerline{\includegraphics[width=5 cm]{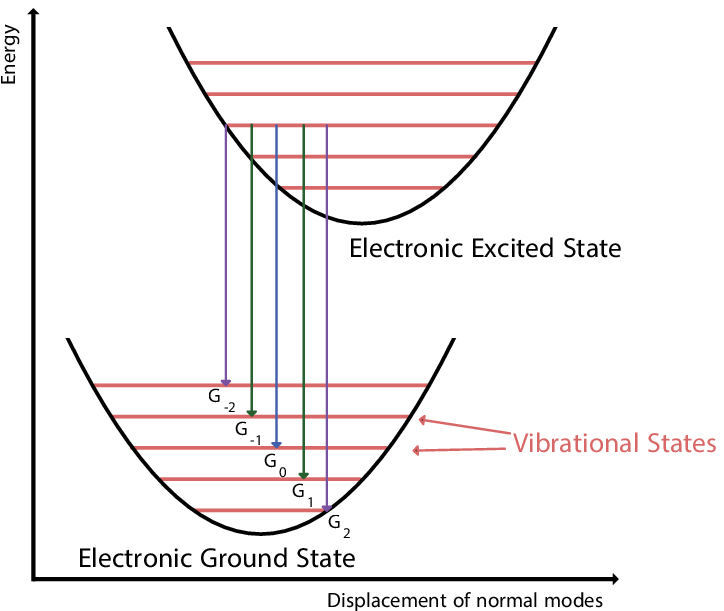}}
\caption{The electron-phonon system for defect center in hBN in the configuration coordinate space. Parabolas represents the harmonic potentials of the lattice in the electronic states.} \label{fig:SFig1}
\end{figure}

A typical emission process from a defect in hBN is shown in Fig.~\ref{fig:SFig1}. The transitions for the zero-phonon line and the phonon sidebands are labelled as $G_0$ and $G_{-2,-1,1,2}$, respectively. 

\section{Temperature dependent optical phonon sidebands}
\begin{figure}[h]
\centerline{\includegraphics[width=9 cm]{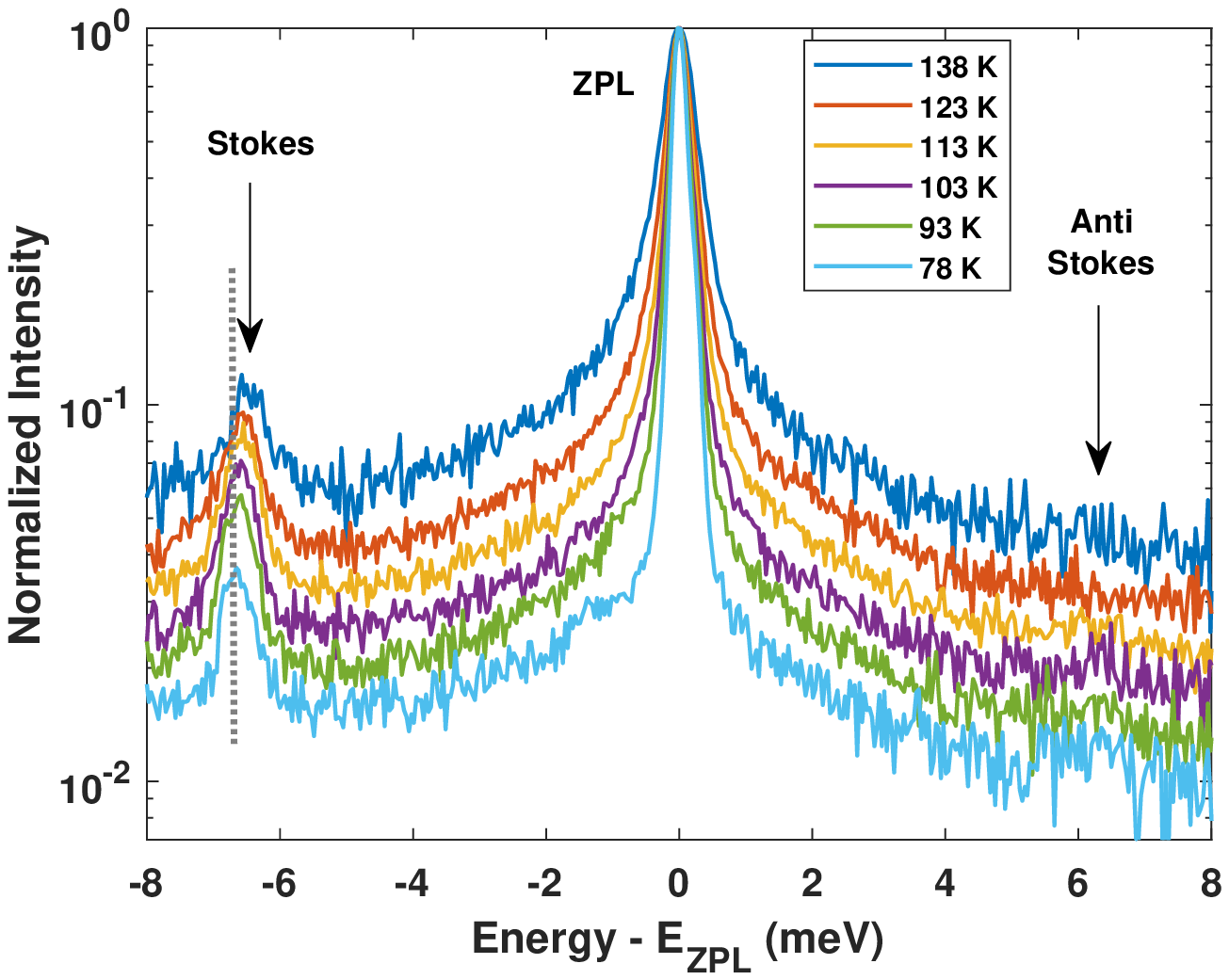}}
\caption{Temperature dependent spectra of the defect taken between 78~K and 138~K. The ZPL energy is centered in each spectrum to observe the temperature dependent detuning between the sidebands and the ZPL energies.} \label{fig:SFig11}
\end{figure}
\section{Temperature dependent broadening and lineshift of ZPL}
The temperature dependent broadening and shift of the ZPL energy can be determined by considering the coupling of low energy LA phonons to the electronic states of the defect. The linear electron-phonon coupling terms introduce a linear temperature dependence to the linewidth as \cite{Stoneham2001}:
\begin{equation}
\Gamma (T) = \left\lbrace8 \ln(2) \sum_\alpha (\hbar \omega_\alpha)^2 S_{\alpha 0} \coth(\hbar \omega_\alpha/2k_BT) \right\rbrace^{1/2} \\
\end{equation}
where $S_{\alpha 0}$ is the Huang-Rhys factor for the phonon with the frequency of $\omega_\alpha$, $k_B$ is the Boltzmann constant. The integral form of the above-equation can be written as:
\begin{equation}
\Gamma (T) = \left\lbrace8 \ln(2) \int_0^{\omega_M} d\omega (\hbar \omega)^2 D(\omega) S(\omega) (2n(\omega)+1) \right\rbrace^{1/2} 
 \label{eq:linG}
\end{equation}
where $S(\omega)$ is Huang-Rhys factor $D(\omega)$ is density of states, $n(\omega)$ is the Bose-Einstein distribution function, and $\omega_M$ is the maximum phonon frequency. Considering the coupling to LA phonons through deformation potential ($V(\omega) \propto \hbar \omega/\sqrt{\hbar \omega }$) and piezoelectric coupling  ($V(\omega) \propto 1/\sqrt{\hbar \omega }$)  with Huang-Rhys parameter $S(\omega)=(V(\omega)/\hbar \omega)^{2}$) and the density of states in 2D ($D(\omega) \propto \omega$), the contribution to the broadening from linear electron-coupling terms becomes $T^{3/2}$ and $T^{1/2}$, respectively.

\begin{figure}[H]
\centerline{\includegraphics[width=10 cm]{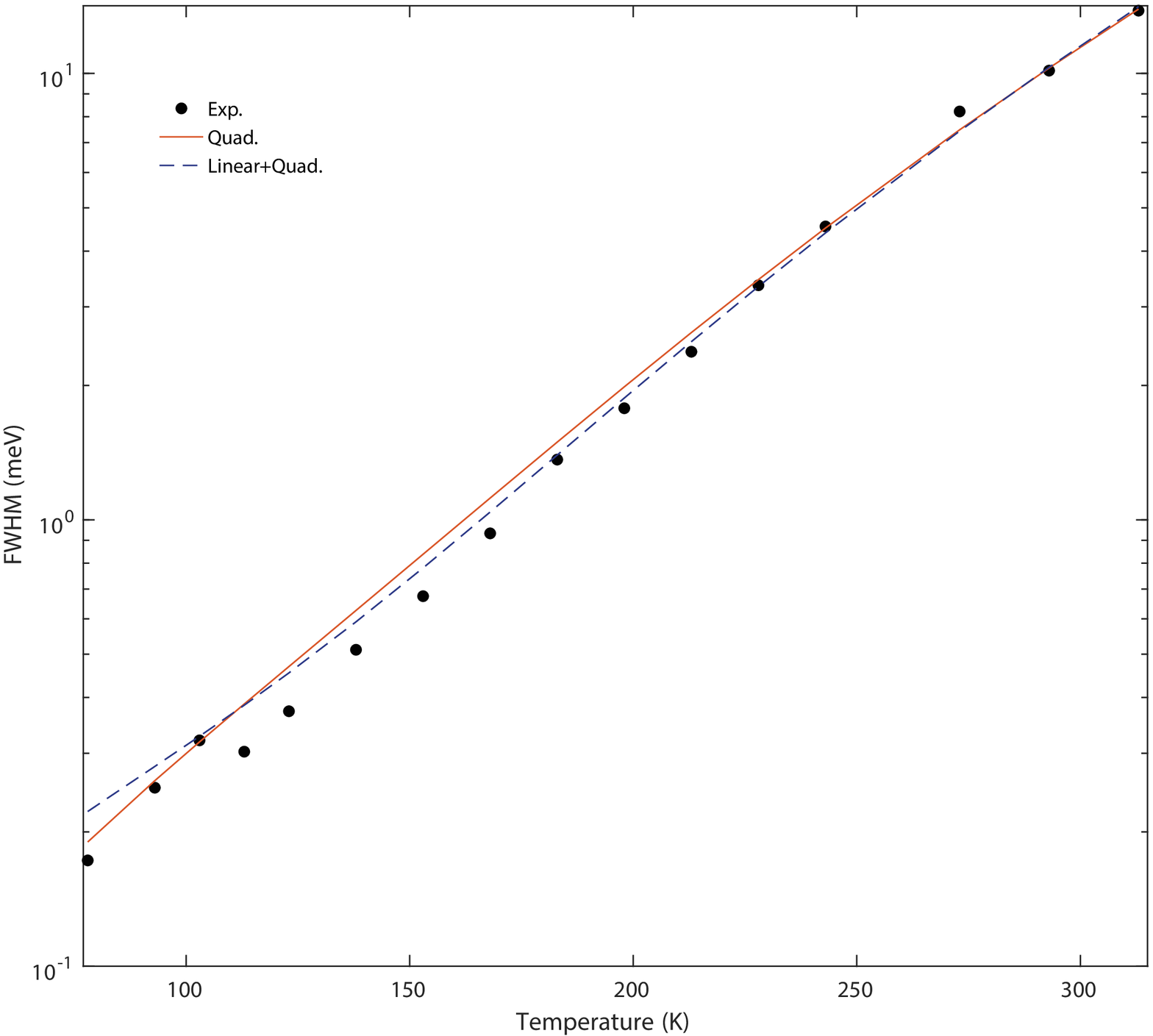}}
\caption{Temperature dependence of the FWHM of the defect center exhibiting a ZPL at 2.2 eV. The experimental data is shown with black dots, while the solid line represents the fit of quadratic coupling terms ($\Gamma(T) = \Gamma_0 + aT + bT^5$) and the dashed line shows the fit  of  linear and quadratic coupling terms ($\Gamma(T)=\Gamma_0+a T^{1/2}+b T^{3/2}+cT+dT^5$) together.}  \label{fig:SFig3}
\end{figure}

The quadratic electron-phonon coupling results in a temperature dependent broadening of the ZPL as \cite{Stoneham2001}:
\begin{equation}
\Gamma(T) \propto \int_0^{\omega_M} d\omega \left\lbrace \sum_q |V_q|^2 \delta(\omega-\omega_q) \right\rbrace^2 n(\omega) \lbrace n(\omega)+1\rbrace 
\label{eq:quadG}
\end{equation}
where $V_q$ is the electron-phonon coupling parameter. Alternatively the integral form of the Equation \ref{eq:quadG} is:
\begin{equation}
\Gamma(T) \propto \int_0^{\omega_M} d\omega \left\lbrace V^2(\omega) D(\omega)\right\rbrace^2 n(\omega) \lbrace n(\omega)+1\rbrace
\end{equation}
Considering the coupling to LA phonons through deformation potential with the density of states in 2D, the contribution to the broadening of the ZPL from the quadratic electron-phonon coupling is derived as:
\begin{equation}
\Gamma(T) \propto T^5
\end{equation} 
Additionally, the coupling of LA phonons through piezoelectric coupling with the density of states in 2D, the contribution to the broadening of the ZPL from the quadratic electron-phonon coupling is derived as:
\begin{equation}
\Gamma(T) \propto T
\end{equation} 
The general temperature dependence of the width from the emission centered  at ZPL becomes:
\begin{equation}
\Gamma(T)=\Gamma_0+a T^{1/2}+b T^{3/2}+c T+d T^5  
\label{eq:gamma}
\end{equation}
However,  second and  third terms approaches to a constant value at low temperatures. In Figure  \ref{fig:SFig3},  fits to the experimental FWHM data with the only quadratic electron-phonon coupling terms and linear plus quadratic coupling terms are shown. Both models represent the experimental data quite efficiently. Therefore, temperature dependence of the FWHM is modelled by considering only the quadratic electron-phonon coupling terms.

The lowest non-vanishing contribution from the electron-phonon interaction to the lineshift is the quadratic coupling which is given as \cite{Stoneham2001}:
\begin{equation}
\Lambda(T) \propto \int_0^{\omega_M} d\omega \left\lbrace \sum_q |V_q|^2 \delta(\omega-\omega_q) \right\rbrace  \lbrace n(\omega)+1\rbrace 
\label{eq:quadL} 
\end{equation}

\begin{equation}
\Lambda(T) \propto \int_0^{\omega_M} d\omega \left\lbrace V^2(\omega) D(\omega)\right\rbrace  \lbrace n(\omega)+1\rbrace 
\label{eq:quadL1}
\end{equation}

Similarly, considering the coupling to LA phonons through deformation potential and piezoelectric coupling with the density of states in 2D, the temperature dependence of the lineshift becomes $\Lambda(T) =cT+ dT^3$. 

\section{Temperature dependent Lineshape analysis of ZPL}

\begin{figure}[H]
\centerline{\includegraphics[width=14 cm]{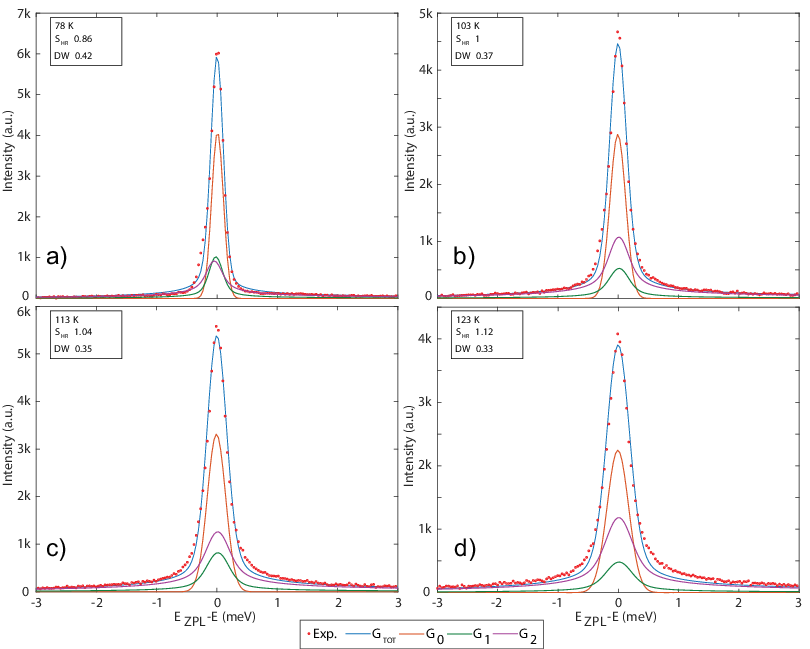}}
\caption{PL emission spectrum from the ZPL (2.2 eV) at various temperatures (a) 78 K, (b) 103 K, (c) 113 K, (d) 123 K are given with symbols. The blue solid lines represent the lineshape calculated from the linear electron coupling theory with the contributions from ZPL (orange line), OPL (green line), and TPL (red line). Calculated Huangh-Rhys and Debye-Waller parameters at relevalent temperature are given in the insets.}  \label{fig:SFig2}
\end{figure}

Figure \ref{fig:SFig2} (a) shows the experimental (symbols) spectral lineshape obtained at 78 K, 103 K, 113 K, 123 K and the result of the fitting with the theoretical lineshape given in the main text including up to two phonon process involving the phonon sidebands from LA phonons coupled to electronic states through deformation potential.

\bibliographystyle{apsrev}
\bibliography{hBN_Paper_Supp_SA}